# Title: Polymer Optical Fiber Fuse

**Authors:** Yosuke Mizuno*, Neisei Hayashi, Hiroki Tanaka, Kentaro Nakamura

**Affiliation:**

Precision and Intelligence Laboratory, Tokyo Institute of Technology, Yokohama, Kanagawa 226-8503, Japan.

* Correspondence to: E-mail: ymizuno@sonic.pi.titech.ac.jp

**Abstract:**

Although high-transmission-capacity optical fibers are in demand, the problem of the fiber fuse phenomenon needs to be resolved to prevent the destruction of fibers. As polymer optical fibers become more prevalent, clarifying their fuse properties has become important. Here, we experimentally demonstrate a fuse propagation velocity of 21.9 mm/s, which is 1–2 orders of magnitude slower than that in standard silica fibers. The achieved threshold power density and proportionality constant between the propagation velocity and the power density are respectively 1/186 of and 16.8 times the values for silica fibers. An oscillatory continuous curve instead of periodic voids is formed after the passage of the fuse. An easy fuse termination method is presented herein, along with its potential plasma applications.

**Main Text:**

Just over a quarter century ago, Kashyap and Blow (*1,2*) published their influential paper on the observation of the optical fiber fuse phenomenon: the continuous self-destruction of a fiber by propagating light. High-power light propagating through the fiber results in local heating and the creation of a high-density plasma or optical discharge that is then captured in the fiber core and travels back along the fiber toward the light source, consuming the light energy and leaving a train of voids (*3*). While fiber fuse propagation is stunningly beautiful (*4*) and may potentially be useful for generating microplasmas at a distance (*5, 6*), the fiber cannot be used after the passage of the fuse. This effect, along with the Shannon limit (*7–9*), nonlinear effects (*10–13*), and the optical amplifier bandwidth (*11*), is now regarded as a critical factor limiting the maximal optical power that can be delivered (*14, 15*). The fuse properties must be well characterized so that all possible measures are taken to avoid the creation of a fiber fuse.

The fuse properties in various glass fibers including standard silica-based single-mode fibers (SMFs) (*1–4, 16, 17*), microstructured fibers (*18*), fluoride fibers (*19*), chalcogenide fibers (*19, 20*), polarization-maintaining fibers (*21*), erbium-doped fibers (*22*), photonic crystal fibers (*23, 24*), and hole-assisted fibers (*25*) are well documented. The fiber fuse is reported to be typically induced at an input optical power of one to several watts (one to several megawatts per square centimeter) and to have a propagation velocity of one to several meters per second. These properties differ according to the type of glass fiber; the threshold power, for instance, is reported to be much higher in photonic crystal and hole-assisted fibers than in silica SMFs (*24, 25*), and nonlinear saturation of the fuse velocity has been observed in erbium-doped fibers (*22*). To date, reports detailing similar properties of non-glass fibers such as polymer optical fibers (POFs) have not been published. Several special POFs with relatively low propagation losses and broadband transmission capabilities have recently become commercially available (*26, 27*) and extensive studies have investigated the implementation of POF-based high-capacity communication systems (*27*) and possible engineering applications of nonlinear effects in POFs (*28, 29*). Therefore, there is a pressing need to clarify the fuse properties of POFs.

We demonstrate here a fiber fuse in a POF that propagates at a velocity of 21.9 mm/s,

which is one to two orders of magnitude slower than that in standard silica fibers. The threshold power density of 6.61 kW/cm$^2$ and the coefficient of proportionality between the propagation velocity and the power density of 1585 mm·s$^{-1}$·kW$^{-1}$·cm$^{-2}$ are 1/186 of and 16.8 times the reported values for silica fibers, respectively. After the passage of the fuse, instead of a void train, we find that an oscillatory continuous curve is formed in the fiber. We also show here that the fiber fuse can be easily terminated using a metal ring and outline the potential applications for remote microplasma generation.

The perfluorinated graded-index POF (*26*) employed in the experiment had a propagation loss of approximately 250 dB/km at 1.55 μm (*30*). Figure 1A depicts the experimental setup, in which output light from a 1546-nm laser was amplified using an erbium-doped fiber amplifier (EDFA) and injected into the POF via an SMF/POF butt-coupling (*28*). We confirmed that the fiber fuse can be initiated in the same way as in glass fibers (*1–4*) by external stimuli such as heating, bending, or bringing the fiber output end into contact with an absorbent material. For the demonstration discussed here, we used a roughly polished POF end surface attached to a smooth silica SMF end surface (*30*).

From observations of the propagation of the fuse along the POF (Movie S1 and Fig. 1B), the propagation velocity was calculated to be approximately 24 mm/s, which is extremely slow in comparison to Todoroki's (*4*) result for a silica SMF. The optical power of the propagating light was calculated using the measured power of the injected light and the loss in the POF to be approximately 75 mW, corresponding to a maximal power density of 7.64 kW/cm$^2$ (*30*). A magnified view of the fuse propagation along a straight portion of the POF is shown in Fig. 1C (70.5 mW, 22.8 mm/s).

We found that the fuse propagation velocity in the POF, measured at 1.55 μm, had an almost linear dependence on the maximal power density with a slope of 1585 mm·s$^{-1}$·kW$^{-1}$·cm$^{-2}$ (Fig. 2A). The power density at which the fuse ceased, i.e., the threshold power density, was 6.61 kW/cm$^2$ at a propagation velocity of 21.9 mm/s. Comparing these results with those of silica SMFs (results at 1.48 μm (*16*) and the theoretical line at 1.55 μm (*17*)) revealed that at 1.55 μm the slope in the POF data (corresponding to the efficiency of the velocity control) was 16.8 times as steep as the silica SMF (94.1 mm·s$^{-1}$·kW$^{-1}$·cm$^{-2}$), and the threshold power density of the POF was 186 times lower than the silica SMF (1.23 MW/cm$^2$). The minimal propagation velocity achieved at 1.55 μm was 5.7 times as low as that theoretically predicted in a silica SMF (125 mm/s).

Digital micrographs taken after the passage of the fuse disclose the extent of the damage to the fiber. The fuse was initially triggered by exploiting the rough surface at the end of the POF (Fig. 3A) and was verified to be induced at the center of the core, which supports the assumption in our calculation that the maximal power density in the fiber cross section affects the fuse induction and can be used to determine the threshold power density (*30*). The passage of the fuse (Fig. 3B) appeared as a continuous black carbonized curve that oscillated periodically along the length of the POF, which is considerably different from the bullet-shape voids observed in glass SMFs. The oscillation period was approximately 1300 μm, which is in general agreement with the theoretical oscillation period of the ray (*30*). Figure 3C shows the position where the fuse ceased after the incident optical power was reduced to that below the threshold; since the fuse remained at this point for several seconds, it melted a relatively large area of the POF, which resulted in the observed bending.

Optical propagation loss in the POF after the passage of the fuse was measured for incremental cutbacks from 30 cm to 20 cm (Fig. 3D) and a fixed input power of 10 dBm (10 mW) at 1.55 μm. A loss of 1.42 dB/cm indicates that, unlike silica SMFs (*1–4*), light can propagate through the POF for several tens of centimeters after the passage of the fuse. We believe this is because undamaged regions remain in the core and cladding layers as these diameters are relatively large. As this loss is somewhat significant for communication applications, once the fuse is induced, it is crucial to stop the propagation as soon as possible.

Several methods for terminating fiber fuses have been developed for glass fibers (*24, 31–33*), which are in principle also applicable to POFs. One method is to thin the outer diameter of the fiber at a certain position while maintaining the core diameter (*32*); this can reduce the internal pressure and arrest the propagating fuse via deformation. In silica SMFs, this structure is fabricated using hydrofluoric acid as an etchant (*32*), but in a POF, chloroform could be used to etch the overcladding layer (*34*). An even easier method, which we present here, is to pressure-bond a small metal ring around the fiber (Fig. S1); this method is only applicable to POFs with an extremely high flexibility (*35*). The propagating fuse is thus terminated by heat absorption and not by deformation. The resulting induced optical loss is negligibly low, and an image of the fuse termination at the position of the ring (Fig. 3E) shows that bending did not occur.

A velocity-adjustable, slowly propagating, low-threshold fiber fuse in a POF is, we predict, a good candidate for creating a microplasma at a distance. Conventionally, two techniques based on spatial laser-light focusing and electrical discharge, respectively, have been used for this purpose (*5, 6, 36*). However, in the former method, an extremely high optical power is required and the optical path in air has a low degree of freedom, while in the latter, an extremely high voltage and relatively large electrodes are required. In contrast, the fiber fuse or microplasma in a POF discussed here can be induced remotely in almost any geometry with only a low-power light without any electrodes. Such a fuse could be utilized as a light source for spectroscopy or a trigger source for another plasma. Thus, these results serve not only as a valuable guideline for the development of POF-based high-capacity transmission systems and engineering applications of nonlinear effects in POFs but also as a strong impetus for the on-going research relating to fiber fuses, material sciences, and especially plasmas.

## Acknowledgments:

The authors thank A. Okino (Department of Energy Sciences, Tokyo Institute of Technology) and H. Yoshida (Sekisui Chemical Co. Ltd.) for the helpful discussions, R. Nedelcov (Department of Music, Tokyo University of the Arts) for the English editing, and K. Minakawa and M. Ding (Precision and Intelligence Laboratory, Tokyo Institute of Technology) for the experimental assistance they offered. This work was partially supported by a Grant-in-Aid for Young Scientists (A) (no. 25709032) from the Japan Society for the Promotion of Science (JSPS) and by research grants from the Hattori-Hokokai Foundation, the Mazda Foundation, the JFE 21st Century Foundation, the General Sekiyu Foundation, the Iwatani Naoji Foundation, and the SCAT Foundation. N.H. acknowledges a Grant-in-Aid for JSPS Fellows (no. 25007652).

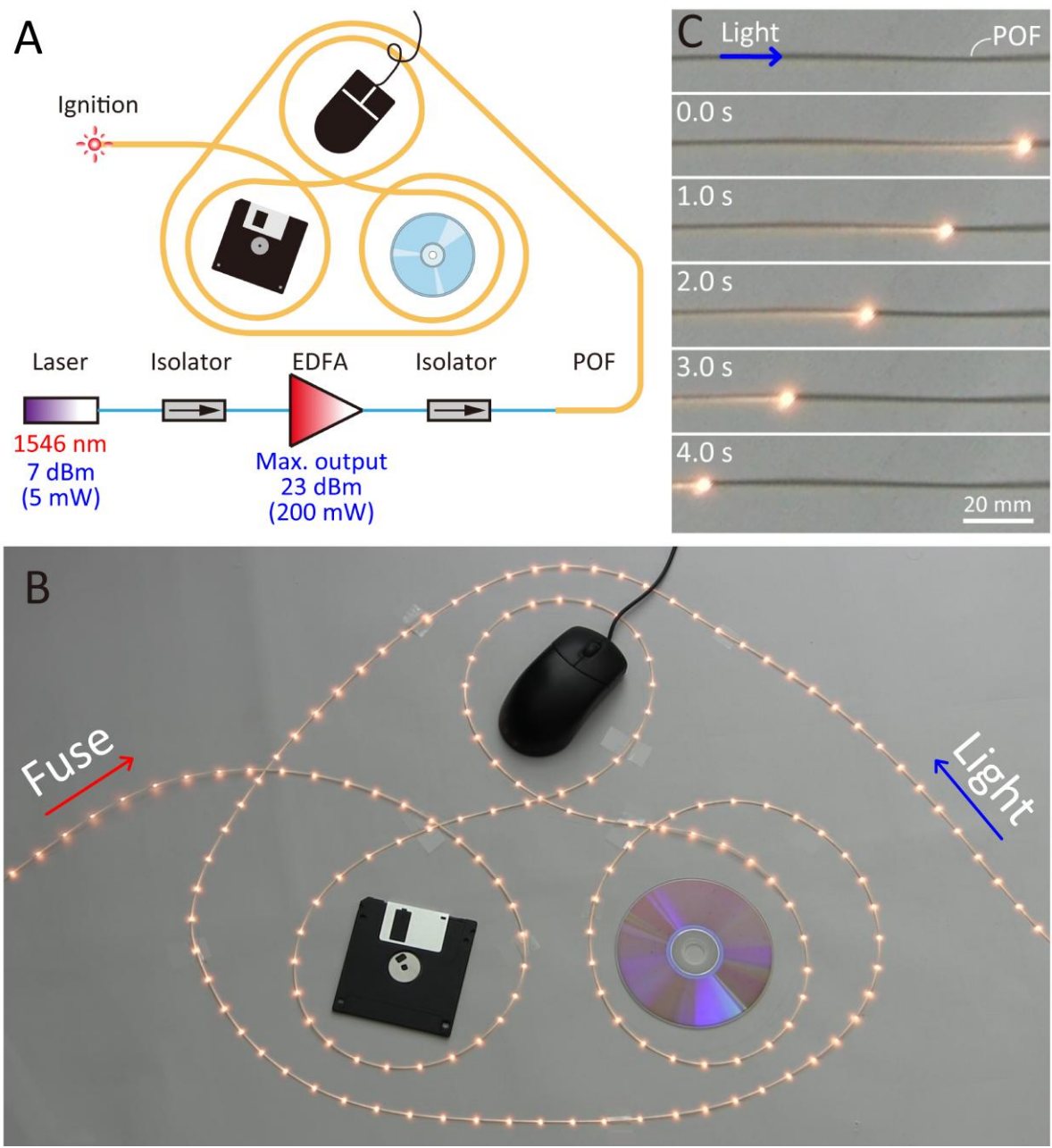

**Fig. 1.** Fiber fuse in a polymer optical fiber (POF). (**A**) Schematic of the experimental setup. The silica SMFs are indicated by blue lines. (**B**) Composite photograph of the fiber fuse propagating along the POF; photographs were taken at 1-second intervals. The light was injected from the right-hand side, while the fuse propagated from the left-hand side. The fiber arrangement was that of Todoroki (*4*) to allow a direct comparison between the POF and silica SMF. (**C**) Magnified view of the propagating fuse. The light was injected from the left-hand side.

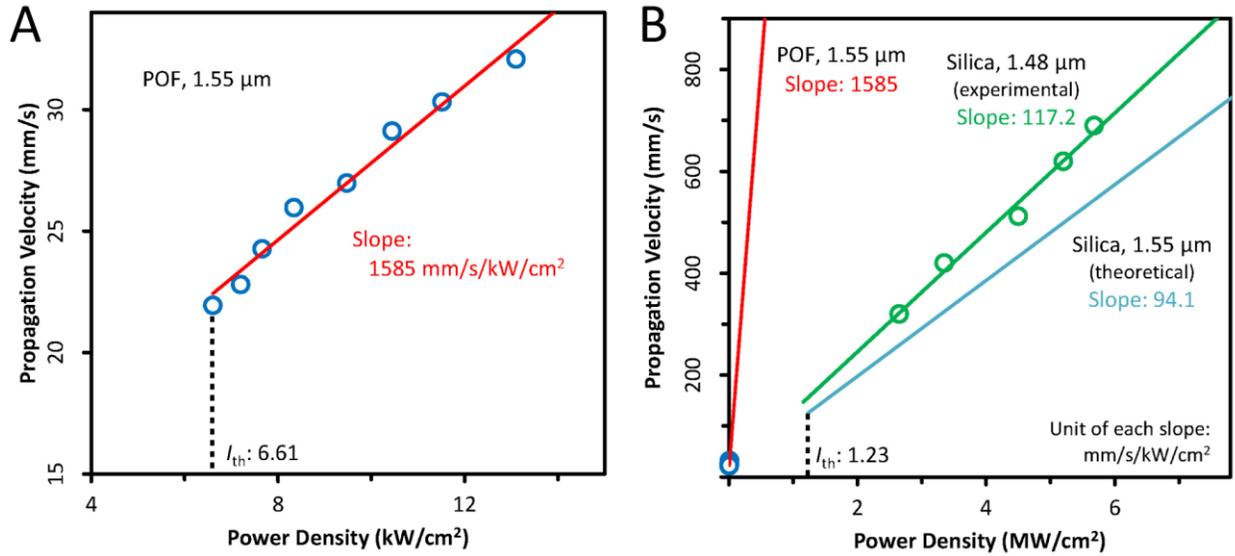

**Fig. 2.** The fiber-fuse properties in a POF and silica SMF. (**A**) Propagation velocity of the fiber fuse in a POF measured at 1.55 μm as a function of the maximum power density in the core. Measured data are shown as blue circles, and the red line is a linear fit. (**B**) Propagation velocities of the fiber fuse as a function of the power density. The measured data for the silica SMF at 1.48 μm (data extracted from the literature (*16*)) are shown as green circles, and the green line is a linear fit; the theoretical threshold power density is 1.16 MW/cm$^2$ (*17*). The blue line is a theoretical prediction for the silica SMF at 1.55 μm (*17*). The data in (**A**) is also reproduced for comparison.

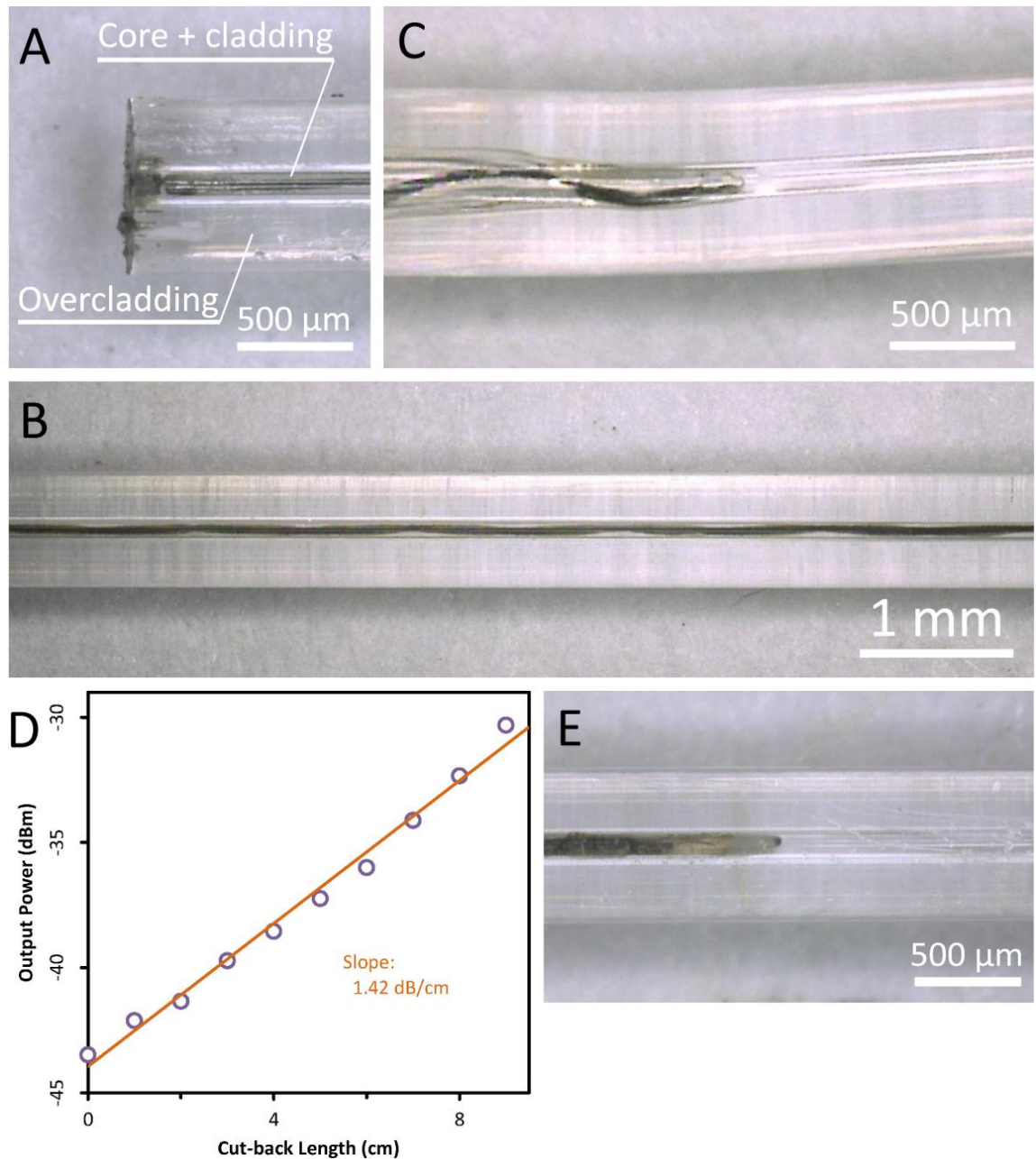

**Fig. 3.** Images of the fiber fuse in the POF. (**A**) Digital micrograph of the POF end at which the fuse was initiated by exploiting the rough surface. (**B**) The path of the fuse in the POF. (**C**) The point at which the fuse was terminated by decreasing the input optical power. (**D**) The dependence of the output power on the cut-back length. The open circles are measured points, and the solid line is a linear fit. (**E**) Image of the fuse termination in the POF at the position of a nickel ring.

# Supplementary Materials:

Materials and Methods

Figure S1

Movie S1

References (*37–40*)

**Materials and Methods:**

The perfluorinated graded-index POF (ID050; Sekisui Chemical Co., Ltd.) (*26*) used in the experiment consists of three layers: the core (50 μm diameter), cladding (100 μm diameter), and overcladding (750 μm diameter). The core and cladding layers are composed of doped and undoped amorphous perfluorinated polymer (polyperfluorobutenylvinyl ether; commercially known as CYTOP), respectively. The refractive index at the center of the core is 1.356, whereas that of the cladding layer is 1.342 (*37*); these values do not depend strongly on the optical wavelength (*38*). The polycarbonate reinforcement overcladding layer reduces microbending losses and increases the load-bearing capability. The fiber is also encased in a white jacket of polyvinyl chloride for further protection. The propagation loss is relatively low (~250 dB/km) even at telecommunication wavelengths, for which inexpensive optical amplifiers such as EDFAs can be used to boost the optical power (note that the standard POF based on polymethyl methacrylate (PMMA) is optimized for visible light transmission, and its propagation loss at 1.55 μm is higher than $1 \times 10^5$ dB/km). This is the main reason for the selection of perfluorinated POFs for the fiber-fuse experiments.

In the experimental setup shown in Fig. 1A in the main text, the 7-dBm (5-mW) output from a distributed-feedback laser diode (NX8562LB; NEC Corp.) with a linewidth of ~1 MHz was amplified by an EDFA (LXM-S-21; Luxpert Technologies Co., Ltd.) to up to 23 dBm (200 mW) and injected into the POF. Two optical isolators were inserted to protect the laser and EDFA from reflected or backscattered light. The end of the silica SMF fitted with an "FC" connector was connected to one end of the POF fitted with an "SC" connector via an FC/SC adaptor. In the demonstration shown in Fig. 1B in the main text, the other end of the POF was polished roughly with 0.5-μm alumina powder and connected to a silica SMF using the same method. By injecting additional light from the silica SMF in the opposite direction, fuse induction can be promoted. Here, the power of the additional light needs to be sufficiently low (16 dBm (40 mW), for example) so that the fuse, once induced, propagates along the POF. All the microscopic images shown in Fig. 3 in the main text were obtained using a digital microscope (VHX-600; Keyence Co., Ltd.). The nickel ring used to terminate the fuse in the POF is shown in Fig. S1.

We can derive an equation for the maximal power density $I$ in the core when light with a certain power $P$ is injected into the graded-index POF. Suppose that the refractive index profile of the core is given by (*26*)

$$n(r) = n_1 \left(1 - 2\Delta \left(\frac{r}{R}\right)^g \right)^{\frac{1}{2}} \cong n_1 \left(1 - \Delta \left(\frac{r}{R}\right)^g \right), \tag{1}$$

$$\Delta = \frac{n_1^2 - n_2^2}{2n_1^2} \cong \frac{n_1 - n_2}{n_1}, \tag{2}$$

where $n_1$ is the refractive index at the center of the core, $n_2$ is the refractive index of the cladding layer, $R$ is the core radius, $g$ is the refractive index profile coefficient, and $\Delta$ is the relative index difference. Under the assumption that all modes propagate with equal attenuation without coupling, the optical power profile is given, in the same way as the refractive index profile, by (*39*)

$$p(r) = p(0)\left(1 - \left(\frac{r}{R}\right)^g\right). \tag{3}$$

Consequently, the maximal power density $I$ can be calculated as

$$I = \lim_{r \to 0} \frac{P}{\pi r^2} \frac{\int_0^{2\pi} d\theta \int_0^r \left(1 - \left(\frac{r}{R}\right)^g\right) r dr}{\int_0^{2\pi} d\theta \int_0^R \left(1 - \left(\frac{r}{R}\right)^g\right) r dr}$$

$$= \frac{P}{\pi R^2} \frac{g+2}{g}. \tag{4a}$$

By assuming $g \cong 2$ in the graded-index POF (*26*), Eq. (4a) can be further simplified as

$$I = \frac{2P}{\pi R^2}, \tag{4b}$$

which indicates that the maximal power density in the graded-index POF with an incident power $P$ is equal to the average power density in a step-index POF of the same core diameter with twice the incident power. For instance, for $P = 75$ mW and $R = 25$ μm, $I$ is calculated to be 7.64 kW/cm$^2$.

In addition, the oscillation period $L$ of the ray in the graded-index POF with $g = 2$ is given by (*40*)

$$L = \pi R \sqrt{\frac{2}{\Delta}}. \tag{5}$$

Using $\Delta = 0.0103$ ($n_1 = 1.356$, $n_2 = 1.342$ (*37*)) and $R = 25$ μm, we obtain $L = 1093$ μm. The discrepancy from the measured value of ~1300 μm appears to be caused by differences in the assumed and actual material parameters as well as structural variations induced during the POF fabrication process.

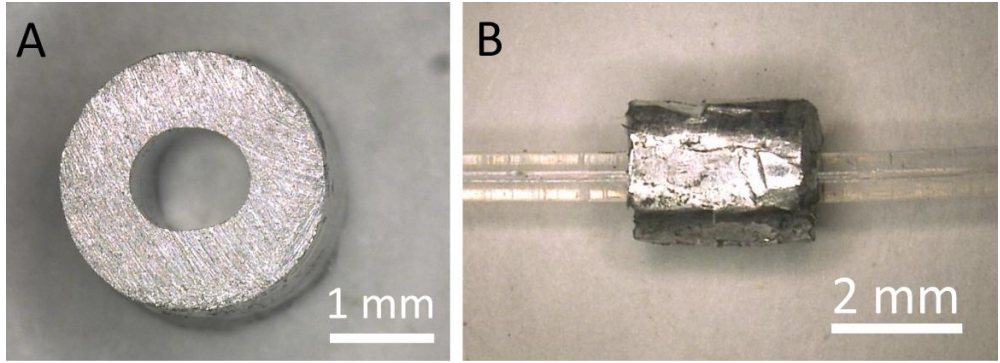

**Fig. S1.** Images of a nickel ring used to terminate the fuse in the POF. (**A**) Photograph of the nickel ring. (**B**) The nickel ring attached to the POF.

**Movie Caption:**

**Movie S1.** The fiber fuse propagating along a POF pumped by ~75-mW, 1.55-μm laser.